\documentclass[conference]{IEEEtran}
\newcommand{\mytitle}{Mutation Analysis with Execution Taints}

\usepackage{textcomp}
\usepackage{tikz}
\usetikzlibrary{chains,shadows.blur,math}
\usepackage{microtype}
\usepackage{comment}
\usepackage{amsmath,amssymb,amsfonts}
\usepackage{footmisc}
\usepackage{algpseudocode}
\usepackage{algorithm}
\usepackage{algorithmicx}
\usepackage{graphicx}
\usepackage{textcomp}
\usepackage{xcolor}
\usepackage{xspace}
\usepackage{boxedminipage}
\usepackage{float}
\usepackage{soul}
\usepackage{alltt}
\usepackage{caption}

\usepackage{mdwtab}
\usepackage{syntax}
\usepackage[procnames]{listings}
\usepackage{lstautogobble}
\usepackage{qtree}

\usepackage{fancyvrb}

\usepackage{hyperref}
\usepackage{cleveref}

\definecolor{rltred}{rgb}{0.5,0,0}
\definecolor{rltgreen}{rgb}{0,0.5,0}
\definecolor{rltblue}{rgb}{0,0,0.5}
\hypersetup{
colorlinks=true,
urlcolor=rltblue, 
linkcolor=rltblue,
citecolor=rltblue,
}

\Crefname{figure}{Fig.}{Figs.}
\crefname{section}{Section}{Sections}
\crefname{subsection}{Section}{Sections}
\Crefname{Algorithm}{Alg.}{Algs.}

\usepackage{caption}
\usepackage{subcaption}

\usepackage{balance}
\usepackage[inline]{enumitem}
\usepackage{tikz}
\usetikzlibrary{shapes.misc, positioning}

\usepackage{booktabs}

\usepackage{float}
\usepackage{listings}
\floatstyle{plain}

\newfloat{lstfloat}{htbp}{lop}
\floatname{lstfloat}{Listing}

\usepackage{color}

\newcommand{\pgmcaesarcypher}{CaesarCypher.py\xspace}
\newcommand{\pgmentropy}{Entropy.py\xspace}
\newcommand{\pgmeuler}{Euler.py\xspace}
\newcommand{\pgmnewton}{Newton.py\xspace}
\newcommand{\pgmprime}{Prime.py\xspace}
\newcommand{\mT}{Traditional\xspace}
\newcommand{\mPS}{Partioned-State\xspace}
\newcommand{\mSS}{Split-Stream\xspace}
\newcommand{\mMS}{Modulo-State\xspace}
\newcommand{\mET}{Exec-Taints\xspace}
\newcommand{\mETNfNm}{Exec-Taints-Nf-Nm\xspace}
\newcommand{\mETNf}{Exec-Taints-Nf\xspace}
\newcommand{\mETNm}{Exec-Taints-Nm\xspace}
\newcommand{\cph}{\emph{competent programmer hypothesis}}
\newcommand{\ce}{\emph{coupling effect}}

\usepackage{hyperref}
\hypersetup{
    pdftitle={\mytitle},    
    pdfauthor={Goerz et al.},     
    colorlinks=true,       
    linkcolor=blue,          
    citecolor=blue,        
    filecolor=cyan,         
    urlcolor=blue        
}

\definecolor{eclipseBlue}{RGB}{42,0.0,255}
\definecolor{eclipseGreen}{RGB}{63,127,95}
\definecolor{eclipsePurple}{RGB}{127,0,85}

\lstdefinestyle{Python}
{
    basicstyle=\footnotesize\ttfamily,
    numberblanklines=false,
    language=python,
    tabsize=2,
    commentstyle=\color{gray},
    keywordstyle=\bfseries\color{eclipsePurple},
    morekeywords={assert},            
    stringstyle=\color{eclipseBlue},
    procnamestyle=\bfseries\color{black},
    procnamekeys={def},
    columns=flexible,
    identifierstyle=
}

\definecolor{codegreen}{rgb}{0,0.6,0}
\definecolor{codegray}{rgb}{0.5,0.5,0.5}
\definecolor{codepurple}{rgb}{0.58,0,0.82}
\definecolor{backcolour}{rgb}{0.95,0.95,0.92}

\lstdefinestyle{mystyle}{
    fancyvrb=true,
    basicstyle=\footnotesize\ttfamily,
    commentstyle=\color{codegray},
    keywordstyle=\color{eclipsePurple},
    escapeinside={(*}{*)},          
    numberstyle=\tiny\color{codegray},
    stringstyle=\color{codepurple},
    breakatwhitespace=false,
    breaklines=true,
    captionpos=b,
    keepspaces=true,
    numbers=right,
    numbersep=5pt,
    procnamestyle=\bfseries\color{black},
    procnamekeys={def},
    morekeywords={where,assert},            
    showspaces=false,
    showstringspaces=false,
    showtabs=false,
    tabsize=2,
    morestring=[b]' 
}
\usepackage{cleveref}
\def\|#1|{\textit{#1}}
\def\<#1>{\texttt{#1}}

\newcounter{todocounter}
\newcommand{\todo}[1]{\textcolor{red}{\stepcounter{todocounter}\footnote[\thetodocounter]{\textbf{TODO }\textit{#1}}}}

\newcommand{\rem}[1]{\textcolor{red}{\textbf{REMOVED }\st{#1}}}
\newcommand{\done}[1]{\textcolor{blue}{\stepcounter{todocounter}\footnote[\thetodocounter]{\textbf{DONE }\textit{#1}}}}

\renewcommand{\todo}[1]{}
\renewcommand{\done}[1]{}
\renewcommand{\rem}[1]{}

\newcommand{\mutation}[1]{$m_{#1}$}
\newcommand{\mutant}[1]{$M_{#1}$}
\newcommand{\mutval}[2]{$M_{#1}:#2$\xspace}
\newcommand{\tmap}[2]{$#1_{\{\text{#2}\}}$}

\ifCLASSINFOpdf
\else
\fi

\begin{document}
\fvset{numbers=left,numbersep=3pt,fontsize=\small,fontfamily=helvetica,frame=lines,resetmargins=true}
%
\title{\mytitle}

\author{
\IEEEauthorblockN{Rahul Gopinath}
\IEEEauthorblockA{University of Sydney\\
Email: rahul.gopinath@sydney.edu.au}
\and
\IEEEauthorblockN{Philipp G\"orz}
\IEEEauthorblockA{CISPA Helmholtz Center\\
for Information Security\\
Email: philipp.goerz@cispa.de}
}


%


\maketitle

\begin{abstract}
Mutation analysis is one of the most effective, but costly means of assessing
the ability of software test suites to prevent bugs.
Traditional mutation
analysis involves producing and evaluating syntactic variants of the original to
check whether the test suite under evaluation is capable of distinguishing
between the variant and the original in terms of behavior.

Evaluating each mutant separately means a large amount of redundant computation,
both between the original program and mutants, and also between different
mutants.

Previous work explored numerous means of removing redundancy. However,
some amount of redundancy has remained especially in the post-mutation phase.

In this paper, we propose \emph{execution taints} -- A novel technique that
repurposes \emph{dynamic data-flow taints} for mutation analysis. Our
technique is the only technique that can remove the redundancy in post-mutation
phase, achieving better efficiency in mutation analysis.
We further leverage memoization to eliminate redundant execution between
program variants.

\end{abstract}


%
\IEEEpeerreviewmaketitle

\section{Introduction}
\label{sec:introduction}
Mutation analysis is the gold standard for evaluating the quality of software
test suites. It is one of the very few techniques that can actually evaluate the
quality of oracles provided by software test suites.
Unfortunately, mutation analysis is also
one of the most costly methods for test suite evaluation.
The problem is that traditional mutation analysis requires generation
and independent evaluation of each syntactic variant (mutant) by the
test suite~\cite{gopinath2017on}.
This means that one needs to execute the entire test suite as many times as the
number of mutants at worst. This can be reduced to some extent by running each
test case only on those mutants whose mutation site is covered by the
test case~\cite{schuler2009javalanche}.
However, this still leaves a large number of mutants to evaluate independently.
For example, \Cref{lst:program} includes four possible mutations:
\mutation{1} in \Cref{line:o4}, \mutation{2} and
\mutation{3} in \Cref{line:o7} and \mutation{4} in \Cref{line:o13}. With
traditional mutation analysis (\Cref{fig:xtraditional}),
this would require five executions including the execution of original to
determine coverage of the test suite, and four from each of the four mutants
\footnote{
We use 
\mutation{x} to denote a mutation where $x$ is the mutation id. The
corresponding mutant is \mutant{x}. We refer to the original as \mutant{0}.
}.

Numerous previous researchers have noticed that traditional mutation analysis
is highly redundant, and significant execution cost savings can be achieved if
one can eliminate some of these redundancies.
Researchers have focused on eliminating these redundancies to make mutation
analysis more efficient.
Just and Fraser~\cite{just2014efficient} showed that not all the mutants need
to be executed independently. Some mutants result in the same value, and hence
the same infected state, and only some of these are propagated.
Hence, one can \emph{partition} states resulting from mutations, and only those
resulting in unique propagated states need to be evaluated independently.
For example,
in \Cref{lst:program}, when \<INPUT=1>, execution of \mutation{1} results in the
same state as the original. Hence, the \emph{partitioned state} execution
(\Cref{fig:xpartition}) skips the execution of \mutant{1}.

In split-stream execution~\cite{king1991fortran,gopinath2016topsy,tokumoto2016muvm},
a mutant is forked when the mutation is executed at runtime, which shares
the execution path until the execution of the mutation.
For example, when one executes the program (\Cref{lst:program}) with say \<INPUT=1>,
at each mutation, a new process is spawned. This results in a total of four
new processes at the end (\Cref{fig:xsplitstream}), but with the initial
execution until the mutation
shared with the original.

The \emph{equivalence-modulo state} approach~\cite{wang2017faster} is the latest
such technique, and it combines the advantages of state-partitioning and
split-stream executions. The idea is to fork only if the mutation leads to a new
unique state. This technique is even more efficient, skipping \mutant{1}
and sharing the execution path between original and also between
mutants (\Cref{fig:xmodulostate}).

\begin{figure}
  \begin{lstlisting}[style=Python, escapechar=|]
def partitioned_process(a: int) -> int:
  i:int = 0
  res: int = 0
  a = a + 1 # m1: a << 1                 |\label{line:o4}|
  counts = get_counts()                  |\label{line:o5}|
  while i < counts:
    a = a/2 # m2: a + 2 m3: a * 2        |\label{line:o7}|
    res += process(i)        |\label{line:o8}|
    i += 1
  return res


def process(i: int):
    if i < 0: # m4: i < 1        |\label{line:o13}|
      return 0
    else:
      return time_consuming(i)


def test_process():
    assert partitioned_process(INPUT) == OUTPUT
        
\end{lstlisting}
  \caption{The \<partitioned\_process()> program}
\label{lst:program}
\end{figure}

\tikzmath{
\yMid = 1;
\yLast = 0;
\nSize=0.3;
\yMidT= \nSize;
}
\tikzset{
pics/mygrid/.style n args={0}{code={
\draw[thick,white,->] (0,0) -- (5,0);
\draw[thick,white,->] (0,0) -- (0,10);
}},
pics/singlestepF/.style n args={5}{code={
\tikzmath{\X=#1;\Y=#2;\T=#3;\yMidF=#4;\yMidT=#5;}
\node[draw,circle,fill=gray!20,inner sep=0pt,minimum size=14pt] (\X.\Y) at (\X,\Y){\T};
}},%
pics/singlestep/.style n args={5}{code={
\tikzmath{\X=#1;\Y=#2;\T=#3;\yMidF=#4;\yMidT=#5;}
\draw [thick,-latex] (\X,\Y+\yMidF) -- (\X,\Y+\yMidT);
\node[draw,circle,inner sep=0pt,fill=white,minimum size=14pt] (\X.\Y) at (\X,\Y){\T};
}},%
pics/singleline/.style n args={4}{code={
\tikzmath{
\nDist=#3;
\yA=#2;
\yB=\yA-\nDist;
\yC=\yB-\nDist;
\yD=\yC-\nDist;
\yE=\yD-\nDist;
\yF=\yE-\nDist;
\yG=\yF-\nDist;
\yH=\yG-\nDist;
\yI=\yH-\nDist;
\yJ=\yI-\nDist;
\yMidF = \nDist-\nSize;
\xP=#1;
}
\draw pic {singlestep={\xP \yA 1 \yMid \yMidT}};
\draw pic {singlestep={\xP \yB 4 \yMidF \yMidT}};
\draw pic {singlestep={\xP \yC 7 \yMidF \yMidT}};
\draw pic {singlestep={\xP \yD {13} \yMidF \yMidT}};
\draw pic {singlestep={\xP \yE 8 \yMidF \yMidT}};
\draw pic {singlestep={\xP \yF 7 \yMidF \yMidT}};
\draw pic {singlestep={\xP \yG {13} \yMidF \yMidT}};
\draw pic {singlestep={\xP \yH 8 \yMidF \yMidT}};
\draw pic {singlestep={\xP \yI {19} \yMidF \yMidT}};
\draw[thick,-latex] (\xP,\yJ+\yMidF) -- (\xP,\yLast) node [below=1pt] {#4};
}},
pics/traditionalM/.style n args={3}{code={
\tikzmath{
\nDist=#3;
\yA=#2;
\xP=#1;
}
\draw pic {singleline={\xP \yA \nDist {M0}}};
\draw pic {singleline={{\xP+1} \yA \nDist {M1}}};
\draw pic {singleline={{\xP+2} \yA \nDist {M2}}};
\draw pic {singleline={{\xP+3} \yA \nDist {M3}}};
\draw pic {singleline={{\xP+4} \yA \nDist {M4}}};
}},
pics/partitionM/.style n args={3}{code={
\tikzmath{
\nDist=#3;
\yA=#2;
\xP=#1;
}
\draw pic {singleline={\xP \yA \nDist {M0}}};
\draw pic {singleline={{\xP+1} \yA \nDist {M2}}};
\draw pic {singleline={{\xP+2} \yA \nDist {M3}}};
\draw pic {singleline={{\xP+3} \yA \nDist {M4}}};
}},
pics/splitstreamM/.style n args={3}{code={
\tikzmath{
\nDist=#3;
\yA=#2;
\yB=\yA-\nDist;
\yC=\yB-\nDist;
\yD=\yC-\nDist;
\yE=\yD-\nDist;
\yF=\yE-\nDist;
\yG=\yF-\nDist;
\yH=\yG-\nDist;
\yI=\yH-\nDist;
\yJ=\yI-\nDist;
\yMidF = \nDist-\nSize;
\xP=#1;
}
\draw pic {singlestep={\xP \yA 1 \yMid \yMidT}};
\draw pic {singlestep={\xP \yB 4 \yMidF \yMidT}};
\draw pic {singlestep={\xP \yC 7 \yMidF \yMidT}};
\draw pic {singlestep={\xP \yD {13} \yMidF \yMidT}};
\draw pic {singlestep={\xP \yE 8 \yMidF \yMidT}};

\draw pic {singlestep={\xP \yF 7 \yMidF \yMidT}};
\draw pic {singlestep={\xP \yG {13} \yMidF \yMidT}};
\draw pic {singlestep={\xP \yH 8 \yMidF \yMidT}};

\draw pic {singlestep={\xP \yI {19} \yMidF \yMidT}};

\draw[thick,-latex] (\xP,\yJ+\yMidF) -- (\xP,\yLast) node [below=1pt] {M0};

\draw pic {singlestepF={{\xP+4} \yB 4 \yMidF \yMidT}};
\draw pic {singlestep={{\xP+4} \yC 7 \yMidF \yMidT}};
\draw pic {singlestep={{\xP+4} \yD {13} \yMidF \yMidT}};
\draw pic {singlestep={{\xP+4} \yE 8 \yMidF \yMidT}};

\draw pic {singlestep={{\xP+4} \yF 7 \yMidF \yMidT}};
\draw pic {singlestep={{\xP+4} \yG {13} \yMidF \yMidT}};
\draw pic {singlestep={{\xP+4} \yH 8 \yMidF \yMidT}};

\draw pic {singlestep={{\xP+4} \yI {19} \yMidF \yMidT}};

\draw[thick,-latex] ({\xP+4},\yJ+\yMidF) -- ({\xP+4},\yLast) node [below=1pt] {M4};

\draw [thick,-] (\xP+\nSize,\yB) -- (\xP+4-\nSize,\yB);


\draw pic {singlestepF={{\xP+2} \yC 7 \yMidF \yMidT}};
\draw pic {singlestep={{\xP+2} \yD {13} \yMidF \yMidT}};
\draw pic {singlestep={{\xP+2} \yE 8 \yMidF \yMidT}};

\draw pic {singlestep={{\xP+2} \yF 7 \yMidF \yMidT}};
\draw pic {singlestep={{\xP+2} \yG {13} \yMidF \yMidT}};
\draw pic {singlestep={{\xP+2} \yH 8 \yMidF \yMidT}};

\draw pic {singlestep={{\xP+2} \yI {19} \yMidF \yMidT}};

\draw[thick,-latex] ({\xP+2},\yJ+\yMidF) -- ({\xP+2},\yLast) node [below=1pt] {M2};

\draw [thick,-] (\xP+\nSize,\yC) -- (\xP+2-\nSize,\yC);

\draw pic {singlestepF={{\xP+3} \yC 7 \yMidF \yMidT}};
\draw pic {singlestep={{\xP+3} \yD {13} \yMidF \yMidT}};
\draw pic {singlestep={{\xP+3} \yE 8 \yMidF \yMidT}};

\draw pic {singlestep={{\xP+3} \yF 7 \yMidF \yMidT}};
\draw pic {singlestep={{\xP+3} \yG {13} \yMidF \yMidT}};
\draw pic {singlestep={{\xP+3} \yH 8 \yMidF \yMidT}};

\draw pic {singlestep={{\xP+3} \yI {19} \yMidF \yMidT}};

\draw[thick,-latex] ({\xP+3},\yJ+\yMidF) -- ({\xP+3},\yLast) node [below=1pt] {M3};

\draw [thick,-] (\xP+2+\nSize,\yC) -- (\xP+3-\nSize,\yC);

\draw pic {singlestepF={{\xP+1} \yD {13} \yMidF \yMidT}};
\draw pic {singlestep={{\xP+1} \yE 8 \yMidF \yMidT}};

\draw pic {singlestep={{\xP+1} \yF 7 \yMidF \yMidT}};
\draw pic {singlestep={{\xP+1} \yG {13} \yMidF \yMidT}};
\draw pic {singlestep={{\xP+1} \yH 8 \yMidF \yMidT}};

\draw pic {singlestep={{\xP+1} \yI {19} \yMidF \yMidT}};

\draw[thick,-latex] ({\xP+1},\yJ+\yMidF) -- ({\xP+1},\yLast) node [below=1pt] {M1};

\draw [thick,-] (\xP+\nSize,\yD) -- (\xP+1-\nSize,\yD);
}},
pics/modulostateM/.style n args={3}{code={
\tikzmath{
\nDist=#3;
\yA=#2;
\yB=\yA-\nDist;
\yC=\yB-\nDist;
\yD=\yC-\nDist;
\yE=\yD-\nDist;
\yF=\yE-\nDist;
\yG=\yF-\nDist;
\yH=\yG-\nDist;
\yI=\yH-\nDist;
\yJ=\yI-\nDist;
\yMidF = \nDist-\nSize;
\xP=#1;
}
\draw pic {singlestep={\xP \yA 1 \yMid \yMidT}};
\draw pic {singlestep={\xP \yB 4 \yMidF \yMidT}};
\draw pic {singlestep={\xP \yC 7 \yMidF \yMidT}};
\draw pic {singlestep={\xP \yD {13} \yMidF \yMidT}};
\draw pic {singlestep={\xP \yE 8 \yMidF \yMidT}};

\draw pic {singlestep={\xP \yF 7 \yMidF \yMidT}};
\draw pic {singlestep={\xP \yG {13} \yMidF \yMidT}};
\draw pic {singlestep={\xP \yH 8 \yMidF \yMidT}};

\draw pic {singlestep={\xP \yI {19} \yMidF \yMidT}};

\draw[thick,-latex] (\xP,\yJ+\yMidF) -- (\xP,\yLast) node [below=1pt] {M0};


\draw pic {singlestepF={{\xP+2} \yC 7 \yMidF \yMidT}};
\draw pic {singlestep={{\xP+2} \yD {13} \yMidF \yMidT}};
\draw pic {singlestep={{\xP+2} \yE 8 \yMidF \yMidT}};

\draw pic {singlestep={{\xP+2} \yF 7 \yMidF \yMidT}};
\draw pic {singlestep={{\xP+2} \yG {13} \yMidF \yMidT}};
\draw pic {singlestep={{\xP+2} \yH 8 \yMidF \yMidT}};

\draw pic {singlestep={{\xP+2} \yI {19} \yMidF \yMidT}};

\draw[thick,-latex] ({\xP+2},\yJ+\yMidF) -- ({\xP+2},\yLast) node [below=1pt] {M3};

\draw [thick,-] (\xP+\nSize,\yC) -- (\xP+2-\nSize,\yC);

\draw pic {singlestepF={{\xP+3} \yF 7 \yMidF \yMidT}};
\draw pic {singlestep={{\xP+3} \yG {13} \yMidF \yMidT}};
\draw pic {singlestep={{\xP+3} \yH 8 \yMidF \yMidT}};
\draw pic {singlestep={{\xP+3} \yI {19} \yMidF \yMidT}};

\draw[thick,-latex] ({\xP+3},\yJ+\yMidF) -- ({\xP+3},\yLast) node [below=1pt] {M4};

\draw [thick,-] (\xP+2+\nSize,\yF) -- (\xP+3-\nSize,\yF);

\draw pic {singlestepF={{\xP+1} \yD {13} \yMidF \yMidT}};
\draw pic {singlestep={{\xP+1} \yE 8 \yMidF \yMidT}};
\draw pic {singlestep={{\xP+1} \yF 7 \yMidF \yMidT}};
\draw pic {singlestep={{\xP+1} \yG {13} \yMidF \yMidT}};
\draw pic {singlestep={{\xP+1} \yH 8 \yMidF \yMidT}};
\draw pic {singlestep={{\xP+1} \yI {19} \yMidF \yMidT}};

\draw[thick,-latex] ({\xP+1},\yJ+\yMidF) -- ({\xP+1},\yLast) node [below=1pt] {M2};

\draw [thick,-] (\xP+\nSize,\yD) -- (\xP+1-\nSize,\yD);
}},
pics/taintedM/.style n args={3}{code={
\tikzmath{
\nDist=#3;
\yA=#2;
\yB=\yA-\nDist;
\yC=\yB-\nDist;
\yD=\yC-\nDist;
\yE=\yD-\nDist;
\yF=\yE-\nDist;
\yG=\yF-\nDist;
\yH=\yG-\nDist;
\yI=\yH-\nDist;
\yJ=\yI-\nDist;
\yMidF = \nDist-\nSize;
\xP=#1;
}
\draw pic {singlestep={\xP \yA 1 \yMid \yMidT}};
\draw pic {singlestep={\xP \yB 4 \yMidF \yMidT}};
\draw pic {singlestep={\xP \yC 7 \yMidF \yMidT}};
\draw pic {singlestep={\xP \yD {13} \yMidF \yMidT}};
\draw pic {singlestep={\xP \yE 8 \yMidF \yMidT}};
\draw pic {singlestep={\xP \yF 7 \yMidF \yMidT}};
\draw pic {singlestep={\xP \yG {13} \yMidF \yMidT}};
\draw pic {singlestep={\xP \yH {8} \yMidF \yMidT}};
\draw pic {singlestep={\xP \yI {19} \yMidF \yMidT}};

\draw[thick,-latex] (\xP,\yJ+\yMidF) -- (\xP,\yLast) node [below=1pt] {M*};



\draw pic {singlestepF={{\xP+1} \yD {13} \yMidF \yMidT}};
\draw [thick,-] (\xP+\nSize,\yD) -- (\xP+1-\nSize,\yD);

\draw [thick,-latex] (\xP+1,\yD-\yMidT) --(\xP+1,\yE) -- (\xP+\nSize, \yE);


\draw pic {singlestepF={{\xP+1} \yG {13} \yMidF \yMidT}};
\draw [thick,-] (\xP+\nSize,\yG) -- (\xP+1-\nSize,\yG); 
\draw [thick,-latex] (\xP+1,\yG-\yMidT) --(\xP+1,\yH) -- (\xP+\nSize, \yH);


\draw [thick,-] (\xP+\nSize,\yC) edge [loop right,min distance=10mm,in=-30,out=30] (\xP+\nSize, \yC);

\draw [thick,-] (\xP+\nSize,\yF) edge [loop right,min distance=10mm,in=-30,out=30] (\xP+\nSize, \yF);
}}
}

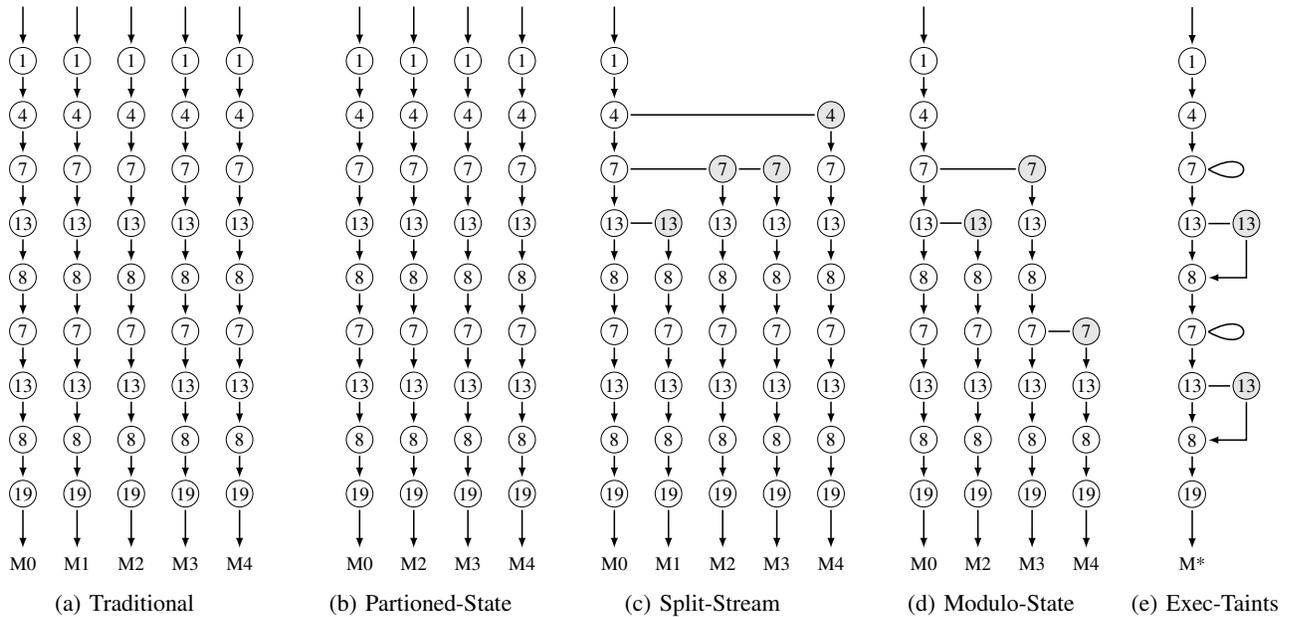
\begin{figure*}[h]
  \centering
\begin{subfigure}[b]{0.24\textwidth}
\scalebox{0.72}{
\begin{tikzpicture}[auto]
\draw pic {mygrid={}};
\draw pic {traditionalM={1 9 1}};
\end{tikzpicture}
}
\caption{\mT}
\label{fig:xtraditional}
\end{subfigure}
\begin{subfigure}[b]{0.18\textwidth}
\scalebox{0.72}{
\begin{tikzpicture}[auto]
\draw pic {mygrid={}};
\draw pic {partitionM={1 9 1}};
\end{tikzpicture}
}
\caption{\mPS}
\label{fig:xpartition}
\end{subfigure}
\begin{subfigure}[b]{0.22\textwidth}
\scalebox{0.72}{
\begin{tikzpicture}[auto]
\draw pic {mygrid={}};
\draw pic {splitstreamM={1 9 1}};
\end{tikzpicture}
}
\caption{\mSS}
\label{fig:xsplitstream}
\end{subfigure}
\begin{subfigure}[b]{0.19\textwidth}
\scalebox{0.72}{
\begin{tikzpicture}[auto]
\draw pic {mygrid={}};
\draw pic {modulostateM={1 9 1}};
\end{tikzpicture}
}
\caption{\mMS}
\label{fig:xmodulostate}
\end{subfigure}
\begin{subfigure}[b]{0.11\textwidth}
\scalebox{0.72}{
\begin{tikzpicture}[auto]
\draw pic {mygrid={}};
\draw pic {taintedM={1 9 1}};
\end{tikzpicture}
}
\caption{\mET}
\label{fig:xtainted}
\end{subfigure}
\caption{The execution flows in the considered mutation analysis algorithms. The numbers correspond to
  line numbers in \Cref{lst:program}. Horizontal lines represent split execution,
  and the loops represent operations within execution taints. \mutant{0} is the
  original execution.}
\label{fig:comparison}
\end{figure*}
That is, the current state of the art~\cite{wang2017faster} is very efficient in
eliminating the causes of redundant execution until the point where mutations
cause a divergence in state. However, the biggest cause of redundancy of
execution is not the pre-divergence execution. There is no solution yet for
avoiding redundant execution in the post-divergence execution phase when a forked
execution restarts following the execution of the original.
For example, the \emph{modulo-state} execution (\Cref{fig:xmodulostate}) forks
a new mutant process from the original after executing \Cref{line:o7} and this
mutant process forks again in the second loop when \mutant{3} and \mutant{4}
diverge. However, in each of these processes, the function \<process(i)> is
called with exactly the same arguments because \<i> is never mutated in these
mutants.

In this paper, we propose mutation analysis with execution taints as a way to
achieve much lower redundancy than previously known methods.
Our proposal is based on three observations:

\begin{description}[leftmargin=0cm]
\item[1.]
The first observation is that many mutants
do not produce a change in the control-flow from the original. For example,
in \Cref{lst:program}, the line \Cref{line:o7} is mutated in mutants
$M2$ and $M3$ resulting in different values for $a$.
Even though the value of $a$ is different
in the original and mutants, since $a$ never influences the control-flow,
the actual execution flow of the original and the mutant is exactly the same.
That is,
except for the values affected by the mutation, the
other steps of execution are the same. Hence, one can reduce the cost
of mutation analysis if such common execution steps can be shared between
such mutants and the original program.

\item[2.]
Our second observation is that for such mutations that does not change control
flow, one can consider the mutation to have introduced a unique \emph{taint} in
the computation, and this taint can be propagated dynamically to any values
using the results of the computation using standard taint propagation through
dynamic data-flow\footnote{
We note that tainting values is a remarkably light weight operation in many
languages. For example, many languages~\cite{perlsec,livshits2012dynamic,haldar2005dynamic}
already has runtime support for tainting, which can be repurposed for execution
taints.}. Such taints can be propagated using standard techniques in most
languages even if the language runtime does not have direct support for taint
propagation~\cite{conti2010taint,fuzzingbook2021:InformationFlow}.
We do not simply propagate a taint flag.
As long as there is no control-flow divergence, we propagate the
results of computations involving mutations in the same taint.
We call such
taints \emph{value taints} of corresponding values.
The computations involving such value taints
are called \emph{execution taints} of the main computation.

\item[3.]
Our third observation is that even when there is a control-flow divergence, it
is possible to limit the impact of diverged control-flow to the function
containing the mutation using \emph{merge-back} of execution and judicious
use of \emph{memoization}.
\end{description}

Using these three observations, our technique (\Cref{fig:xtainted}) can perform
extremely efficient mutation analysis limiting both pre and post mutation
redundancies .

\subsection*{\textbf{Contributions}}
\begin{itemize}
\item We propose a complete framework for avoiding redundancy in mutant execution
  including both pre-divergence and post-divergence stages.
\item Our \emph{execution taints} is a lightweight mechanism that repurposes
  dynamic data-flow taints for mutation analysis.
\item We propose a novel memoization strategy for sharing computational results
  across mutants.
\item We propose a comprehensive solution to the problem of sharing single-use
  resources by mutants during forking.
\end{itemize}

\section{Approach}
We now discuss our approach in detail. There are two cases that we need to
discuss in detail. The first is when there is no control-flow divergence between
the mutant and the original. The essential idea is that values resulting from
mutated expressions are stored as \emph{dynamic taints} of the resulting values.

\subsection{Taint Schemata}
Given a program and test suite that needs to be evaluated, we first transform
the program to a meta-mutant containing all mutants.
For example, below is the code fragment from \Cref{lst:program}.
\begin{lstlisting}[style=Python, escapechar=|,firstnumber=6]
while i < counts:
  a = a/2
  res += process(i)
  i += 1
\end{lstlisting}
This gets transformed into the following meta-mutant\footnote{The chevrons ($<$,$>$) indicate
lambdas, which are also passed the mutation id.}
\begin{lstlisting}[style=Python, escapechar=|,firstnumber=6]
while i < counts:
  a = @T(<M0: a/2>, <M2: a+2>, <M3: a*2>)
  res += process(i)
  i += 1
\end{lstlisting}
where \mutant{0} is the key for original value, \mutant{2} is the key for second
mutant, and \mutant{3} is the key for the third mutant. 
During the execution of the meta-mutant,
the function \<@T> is called and passed the different possible variants of
\Cref{line:o7}. The function \<@T> collects the results of each, and these are
stored in the dynamic taint of \<a> as a key-value map.

\subsection{Execution Taints and Transmission Rules}
\label{sec:execution-taints}
As we specified in \Cref{sec:introduction}, We use \mutation{x} to denote a mutation where
$x$ is the mutation id. The corresponding mutant is \mutant{x}.
We refer to the original as \mutant{0}.

The single taint value corresponding to a single mutant is called an execution
taint of that value for that mutant, and the list of taints from different mutations
that is attached to a value (or equivalently, a variable at a point in
execution) is called the \emph{taint map} for that value.
That is, after the execution of \Cref{line:o7}, if we started with a value
$a=2$, then the resulting $a$ will have a value given by the taint map:
\{\mutval{0}{1}, \mutval{2}{4}, \mutval{3}{4}\}. Here, 4 is an execution taint corresponding
to both \mutant{2} and \mutant{3} attached to the original value. 1 is the execution
taint of the original (\mutant{0}) -- i.e. the non-mutated value.

When two values are used in an operation, the following rules are used for taint
propagation. We use the $A$, $B$, and $C$ for indicating the variables involved
in an expression. We further use $\otimes$ to indicate an operation.

We use $A^{x}$ to indicate the value taint of \mutant{x} stored in $A$.
If $A$ does not have a value taint of \mutant{x}, the original value, that is,
the value taint of \mutant{0} is returned instead.

If a
variable is tainted, it annotated with its taint map. For example,
\tmap{A}{\mutval{0}{A^0},\mutval{1}{A^1}} means that $A$ carries an execution
taint of mutation \mutant{1}.
It is equivalent to \tmap{A}{\mutval{1}{A^1}} eliding the original \mutant{0} value.

That is, given $M^x \subset M$, $M^y \subset M$, and
$M_o$ is an operator mutation in $M$, where $M$ is the set of mutants, after
the execution of
\[
  A_{\{M_i:A^i \; \mid \; M_i \in M^x \}} \; \otimes_{M_o:\odot} \; B_{\{M_j:B^j \; \mid \; M_j \in M^y \}} 
\]
the result will be stored as
\[
  C_{\{M_i:(A^i\otimes{}B^i) \; \mid \; M_i \in (M^x \cup M^y) \}\;\cup\;\{M_o:A\odot{}B\}}
\]

We discuss a few clarifications of this equation next. We elide the braces in
the taint map when there is no ambiguity in the context.

\begin{enumerate}
  \item If neither value has a taint, and the operation is not a mutation,
then the resulting value will not have a mutation execution taint. That is, given the operation
\[
  A \otimes B
\]
where the result is stored in $C$, $C$ will not have a taint.
The above equation is equivalent to
\[
  A_{M_0:A} \otimes B_{M_0:B}
\]
    with $C_{M_0:A\otimes{}B}$ as the result.

\item An operation can also be tainted when it itself is mutated. In that
case, the operation will be associated with a single mutant.
E.g. $\otimes_{M_x:\odot}$ indicates that $\otimes$ was mutated by
\mutation{x} changing $\otimes$ to $\odot$.

  \item If neither value has a taint, but the operation is a mutation, the
    resulting value will have the taint of the mutation as the single element
    in its taint map.
    That is, given the equation
\[
  A \otimes_{M_x:\odot} B
\]
    the result will be stored as $C_{M_x:A\odot{}B}$.
    For example, given $a + b$ as the original expression,
    with $a=1$, $b=1$, and we have mutation \mutation{1} that mutates this expression to
    $a - b$, the resulting value will have
    \{\mutant{0}:1, \mutant{1}:0\} as the taint map.

\item If both values have \mutant{x} in their taint map, and the operation
  is not mutated, then the operation will be performed between the values
  corresponding to \mutant{x} in both values' taint map.
That is, given the equation
\[
  A_{M_x:A^x} \otimes B_{M_x:B^x}
\]
    the result will be stored as \tmap{C}{\mutval{X}{A^x\otimes{}B^x}}.
    For example, given \tmap{a}{\mutval{0}{1},\mutval{1}{2}} and
    \tmap{b}{\mutval{0}{3},\mutval{1}{4}}, $c = a + b$
    will result in \tmap{c}{\mutval{0}{4}, \mutval{1}{6}}.

\item If both values have \mutant{x} in their taint map, and the operation
  is also mutated, then the mutated operation will be performed between the
  values corresponding to \mutant{x} in both values' taint map.
That is, given the equation
\[
  A_{M_x:A^x} \otimes_{M_x:\odot} B_{M_x:B^x}
\]
    the result will be stored as \tmap{C}{\mutval{x}{A^x\odot{}B^x}}

\item If one of the values is missing the taint \mutant{x} in their taint map,
and the other has \mutant{x} in their taint map, the value of \mutant{0} will be
used instead, and the resulting value will be placed as execution taint
corresponding to \mutant{x} in the taint map.
That is, given the equation
\[
  A_{M_x:A^x} \otimes B
\]
the result will be stored as \tmap{C}{\mutval{x}{A^x\otimes{}B^0}}.

\item In the case of multiple execution taints, each taint is evaluated separately.
That is, given the equation
\[
  A_{M_x:A^x, M_y:A^y} \otimes_{M_z:\odot} B_{M_x:B^x, M_y:B^y}
\]
the result will be stored as \tmap{C}{\mutval{x}{A^x\otimes{}B^x}, \mutval{y}{A^y\otimes{}B^y}, \mutval{z}{A\odot{}B}}.
Note that different mutants never interact.
For example, given \tmap{a}{\mutval{0}{1},\mutval{1}{2}} and \tmap{b}{\mutval{0}{3},\mutval{2}{5}}, $c = a + b$
will result in \tmap{c}{\mutval{0}{4},\mutval{1}{5},\mutval{2}{6}}.
\end{enumerate}

For example, the \<a> after \Cref{line:o7} in the first iteration of \mbox{\<partitioned\_process(1)>} is
annotated \tmap{a}{\mutval{0}{1},\mutval{2}{4},\mutval{3}{4}}.
This corresponds to the value of $a=1$ in the
original and $a=4$ in both \mutant{2} and \mutant{3}.
That is, each mutant shares the execution as well as the state until the
mutation is seen, after which it shares the execution and the variables not
tainted in the mutation with the original. 

Such taints are checked during test case assertions. Each single execution
taint that correspond to a mutant is executed separately in each assertion.
Any mutant that fail the assertion is marked as \emph{killed}.

Propagating execution taints works well until the control-flow diverges.
If a tainted value is involved in control-flow, and the control-flow
diverges from the mainline\footnote{We call the execution path of the
original program \emph{mainline} execution path.} the simple data-flow
taint transmission is no longer sufficient for executing mutants that
diverged.
However, for conditionals, (assuming no \<return> statements), the
control-flow merges back to the mainline after the conditional branch finishes
execution. Similarly, for loops, the control-flow merges back to the mainline
execution path after all iterations finish. However, the presence of
\<return> statements can introduce additional complexity.
Hence, we use the procedure boundaries as merge points of diverged
execution.

In this case, we have two choices based on whether the platform
has the ability to \emph{fork} (and is efficient to do so)

\subsection{Diverged Control-Flow When Fork is Available}

Accounting for diverged control-flow when an efficient fork is available is
accomplished as follows. We first wrap every function in the program with
another function (In Python, this is accomplished using a decorator~\cite{hunt2019python}).
Next, we implement the split-stream execution within the wrapped function,
but rather than splitting on encountering a mutation, we split on control-flow
divergence to as many children as there are diverging mutant executions.

That is, as soon as one of the mutant control-flow paths diverge from the
main, the process forks, with the parent continuing the mainline execution,
and the children following the divergent control-flows. Given that this is
a function call made from a wrapper function, the wrapped function returns
in mainline as well as divergent processes to the wrapper function. Within the
wrapper function, the parent process waits for, and gathers all global variable
changes induced by the divergent processes as well as the return value from the
wrapped function\footnote{We call this the \emph{merge-back} of diverging executions
both for forking and non-forking choices.}.
These are updated to the taint maps of variables in the mainline.
The execution continues from the wrapper function returning the
value with the updated taints.

\subsection{Diverged Control-Flow When Fork is Unavailable}

Accounting for diverged control-flow when fork is unavailable or is too
inefficient to use either because the operating system fork is costly or
the process of merge-back from external process is costly. As before, we
start with a function wrapper for each functions.

Within such a wrapper, we first set the execution path to be followed to
\mutant{0}, that is the mainline. Any mutant that does not follow the execution
path of \mutant{0} is marked \emph{wounded}.

Once the function finishes for the given execution path, it returns to the
wrapper function. Within the wrapper, we look at the mutants that are newly
marked as \emph{wounded}, and those that are marked as \emph{active}.
We temporarily remove active mutants from consideration, choose one of the
wounded mutants -- say \mutant{x}. For any such mutant, we can determine if
the wound happened due to a mutation within the function (or any child functions
called from the function) or if the function and child functions were
unmodified because for any active taints, we know the location where the taint
was encountered.

Let us take the case when the function and child functions were unmodified.
In this case, we execute the wrapped function
using untainted parameters with values from execution taints from \mutant{x}.
This gives us the return value from the mutant. The return value is merged
back as execution taint of \mutant{x} on to the mainline. The taints of
any globals modified by the function call are similarly updated. We mark
\mutant{x} as no longer wounded. We pick the next wounded mutant, and continue
the process until there are no more wounded mutants.

If on the other hand, the function or one of the child functions called from
it contained the current mutation, we call the mutated function instead for
that function, and collect taints. The remaining process is the same.

Thus at the end of the function call, we have successfully merged the diverged
control-flows into the system state.

\subsection{Memoization for Sharing Execution Across Mutants}
At this point, we have shown that we can merge back executions from
diverging control-flows at the function boundary where the diverging execution
happened. However, this still leaves some redundancy. Consider
the fragment in \Cref{lst:myfn}.
\begin{figure}[H]
\begin{lstlisting}[language=Python, escapechar=|]
def my_fn(v: int) -> int:
    if v > 0: # m1: v > 1
        r = time_consuming_fn(v)
        return r
    else:
        r = time_consuming_fn(v)
        return -r
\end{lstlisting}
\caption{Memoization Example}
\label{lst:myfn}
\end{figure}
Here, the control-flow diverges at \mutation{1} when \<my_fn(1)> is
executed. However, \<time\_consuming\_fn(v)> is executed in both original
as well as the mutant with exactly the same parameters. Hence we next see
how to eliminate this redundancy.

The basic intuition here is that given that we already wrap each function,
we can memoize the execution results of each function and specific
parameter values. (This database of memoized function calls are stored 
externally for the forking version so that all processes can access it.)

An important constraint for memoization is that we should ensure that it
does not grow out of bounds. Since concern here is ensuring a fast mutation
analysis by sharing similar executions across mutants, we follow the
simple expedient. The memoization cache is maintained only if there are
unmerged mutants. We clear the current cache completely as soon as all
mutants merge. The idea here is that a mutant can make use of the cache
for avoiding redundant execution with the original or another mutant
if and only if there are active unmerged mutants in the first place.
We also ensure that the original process executes first, and the mutant
processes start execution only when the original has returned from the function.
Further, only one mutant process is executing at any time. (If parallel
execution is more important than utilizing memoization, one can start the
mutant-processes in parallel).

There are two cases to consider for updating the memoization cache.
The first is when no mutation is encountered within the function call (or in one
of the child functions called from that function), and the second when a
mutation is encountered within the function call (or its child calls).

New processes update the cache if and only if the \emph{mutation} was not
encountered after forking. This is because we can't share the execution of
a mutated function with any other mutant.
To implement this, we maintain a stack of current function calls. That is,
each entry in the stack is of the form all:(function, parameter values).

Say we have a function call with a number of active
execution taints in the parameters. We execute the mainline, and at the end
of execution of the mainline, we have a subset of these execution taints in the
return value.
If any mutation was encountered (the execution of \<taint\_t()> call), 
we look at the call stack and we produce the following entry for current
call and each parent call --- (mutant, call:(function, parameter values)) = \<mutated>.
This is saved in a mutation cache.


Each execution taint in the return value of the mainline represents a possible
new entry in the memoization cache.
That is, we add a memoization entry with the key
call:(function, parameter values), and the corresponding return value is stored
with this key unless the mutation cache contains the just returned call for this
particular execution taint. If it contains the just returned call, then the
return value is not memoized.

For each child processes, before executing any functions, the memoization is
first checked to see if the call was previously memoized, and the call is not
present in the mutation cache. If so, the result
is retrieved, and used. The memoization cache is updated with the result of
a function if the entry did not previously exist, and the mutation cache does
not contain the current (mutant, call:(function, parameter values)) tuple.

With this technique, we guarantee that any non-mutated function call (that is,
function calls that do not encounter a mutation during the execution of the
current function or its children) within the mainline or the mutants will be
available by other mutants.

If the function call encountered a mutation, then obviously the function being
computed is different, and it is not shareable by other mutants.

\section{Implementation}
We were motivated by a desire to place our ideas before the mutation analysis
community at an early phase so that the community can build on our ideas, and we
may gain valuable early feedback. Hence, we were motivated to quickly flush out
our initial prototype demonstrating the proof of concept that our ideas can
indeed work.

However, this also meant that we chose a number of techniques based on their
simplicity rather than their efficiency. For example, our prototype is
implemented in Python, and uses a library based taint mechanism.

We wanted to compare each of the previous approaches and identify how they
improved over traditional mutation analysis, and previous state of the art.
Hence, we implemented each of the approaches ourselves. These include the
(1) simple traditional approach, (2) the split-stream approach, and (3) the
equivalence modulo state approach. Further, we wanted to evaluate the
effectiveness of the two different kinds of merge back: with forking and
without forking both without and with memoization.

\subsection{Implementation of Taints}

As a prototype, our priority was on evaluating the realizability of our proposal
quickly. Hence, the taint transfer is only implemented for the primitive
types int and float, which were sufficient to completely implement mutation
analysis for our subjects. However, as Conti~\cite{conti2010taint} demonstrates,
this approach can be extended to arbitrary data types in Python.
The primitive types are wrapped by a proxy class, providing a proxy object that
behaves like the original, but with the added capability to propagate taints.
The taints are stored as a Python dictionary with the mutation id as the key
and the value corresponding to the mutation as the value.



\subsection{Mutation Operators}

The following mutation operators are implemented in our mutation framework.
\begin{itemize}
  \item \textbf{Comparison operators.} We mutate each of the following comparison operators \(==, !=, <, <=, >, >=\) to
each other.
\item \textbf{Arithmetic operators.} The following arithmetic operations are mutated into each other
\(+, -, *, \%, <<, >>, |, ^, \&, //\).
\end{itemize}

\subsection{Meta Mutant}
For implementing our technique, we use AST transformation to embed the different
mutations in the same source code. We show the transformation of the program in \Cref{lst:program}
in \Cref{lst:metamutant}.
\begin{figure}[H]
  \begin{lstlisting}[style=Python, escapechar=|]
def partitioned_process(a: int) -> int:
  i:int = 0
  res: int = 0
  a = @T(<M0: a+1>, <M1: a<<1>) |\label{line:m4}|
  counts = $get_counts()
  while @C(i < counts):
    a = @T(<M0: a/2>, <M2: a+2>, <M3: a*2>) |\label{line:m7}|
    res += $process(i)
    i += 1
  return res
$partitioned_process = wrap(partitioned_process)

def process(i: int):
    if @C(@T(<M0: i<0>, <M4: i<1>)): |\label{line:m13}|
      return 0
    else:
      return $time_consuming(i)
$process = wrap(process)

def test_process():
    @assert $partitioned_process(INPUT) == OUTPUT
\end{lstlisting}
  \caption{The meta-mutant of the \<partitioned\_process()> program}
\label{lst:metamutant}
\end{figure}
Each mutation point is replaced by a \<@T()> call with a dictionary as the
parameter. The dictionary contains several key value pairs of the format
$\langle$\<Mi: expr>$\rangle$. Here, the key for the dictionary is \<Mi> which
is the mutation id. The \<expr> is the original or mutated expression wrapped
in \<lambda> that is evaluated in the context of \<Mi>.
Each lambda expression is evaluated to generate the corresponding execution taint.
For the evaluation of a \<lambda> corresponding to a particular mutation, the
corresponding execution taints of that mutation is extracted from the expression.
That is, $\langle$\<M2:a+2>$\rangle$ indicates that we first extract the
execution taint \<M2> in the variable \<a> (if not found, we extract the execution taint \<M0>
(the original value) from \<a>), and use it for computing the expression \<a+2>.
This is then stored under the key \<M2> the execution taints of the resulting expression.
Any uncaught exception during the evaluation of such a lambda function is counted as a
strong kill of that mutant.
%
\subsection{Implementation of Forking}
Forking is performed in the \<@C(expr)> call when evaluating the expression
\<expr> results in different values for the mainline and some execution taint
corresponding to a mutant. After forking, the mainline and mutants that follow
the mainline continues in the parent process while the mutants that diverged
continue in the forked child. The forked process (and the parent) can fork again
if the mutants in that process diverge from each other again.

Each function is wrapped in a wrapper function (the wrapped functions are
indicated by a \$ in their name) that detects a control-flow divergence within
the function. When there is a control-flow divergence (and hence a forking) the
parent process starts executing (the child processes are immediately suspended)
and finishes first, returning to the wrapper function. The wrapper then
starts the child processes one at a time. Each child process exits when it
reaches the wrapper function after communicating the result of the function call
to the parent process. The wrapper function in the parent process receives
these results, updates the taint of the function return and continues.



\section{Evaluation}

For evaluation, we chose the following subjects: (1)~\pgmcaesarcypher, (2)~\pgmentropy, (3)~\pgmeuler, (4)~\pgmnewton, (5)~\pgmprime.
The details are given in \Cref{tbl:subjects}. As our technique is in the
prototype stage, the subjects were chosen for the limited data types, and Python
features that they used. Our subjects are also somewhat small, implementing just
a simple algorithm in each case. Hence, this our evaluation is not a general
empirical validation, but rather a proof of concept that this technique can
work.
\begin{table}
  \centering
\begin{tabular}{|l|r|r|}
  \hline
Subject       & LOC & Mutants \\
\hline
\pgmcaesarcypher & 42  & 55 \\
\pgmentropy      & 19  & 46 \\
\pgmeuler        & 19  & 35 \\
\pgmnewton       & 15  & 39 \\
\pgmprime     & 24  & 58 \\
\hline
\end{tabular}
\caption{Subjects}
\label{tbl:subjects}
\end{table}

\begin{table*}[tb]
\center
\begin{tabular}{|l|r|r|r| r|r|r|r|}
\hline
Program       & \mT & \mSS & \mMS & \mETNfNm & \mETNf & \mETNm & \mET \\
\hline
\pgmcaesarcypher & 30725  & 31517 & 22803  & 2830     & 2830 & 988 & 988 \\
\pgmentropy      & 4974   & 4453  & 4106   & 108      & 108   & 108  & 108  \\
\pgmeuler        & 3786   & 3469  & 2510   & 1192     & 1192  & 140  & 140  \\
\pgmnewton       & 2095   & 1737  & 1239   & 511      & 511   & 103  & 103  \\
\pgmprime        & 5057   & 4533  & 3059   & 2740     & 2020  & 931 & 931 \\
\hline
Mean $T\times$   & 1      & 0.89  & 0.68   & 0.25     & 0.23  & 0.07 & 0.06 \\
\hline
\end{tabular}
\caption{The number of lines of code from the program that is executed by various mutation analysis techniques.}
\label{tbl:evaluationcode}
\end{table*}

\begin{table*}[tb]
\center
\begin{tabular}{|l|r|r|r| r|r|r|r|}
\hline
Program       & \mT & \mSS & \mMS & \mETNfNm & \mETNf & \mETNm & \mET \\
\hline
\pgmcaesarcypher & 0   & 519532 & 1102095 & 4283465 & 7698959 & 2346929 & 3562687\\
\pgmentropy      & 0   & 131935 & 313604 & 255611  & 264010  & 248328 & 256665 \\
\pgmeuler        & 0   & 86980  & 165647 & 1999129 & 2142995 & 216044 & 233252 \\
\pgmnewton       & 0   & 420113 & 783796 & 4954838 & 5002992 & 752975 & 759959 \\
\pgmprime        & 0   & 149927 & 272727 & 2166888 & 1600531 & 688069 & 715081 \\
\hline
\end{tabular}
\caption{The infrastructure code executed to support each technique per program. We note that the infrastructure code is very large due to the prototype nature of our implementation. This can be optimized heavily.}
\label{tbl:evaluationinfra}
\end{table*}

\begin{table*}[tb]
\center
\begin{tabular}{|l|r|r|r| r|r|r|r|}
\hline
Program       & \mT & \mSS & \mMS & \mETNfNm & \mETNf & \mETNm & \mET \\
\hline
\pgmcaesarcypher & 3.38   & 11.50 & 15.75 & 104.73 & 77.06 & 103.06 & 113.75 \\
\pgmentropy      & 2.63   & 5.56 & 8.83 & 3.30 & 3.26 & 3.29 & 3.32 \\
\pgmeuler        & 1.93   & 3.98 & 3.58 & 19.02 & 20.53 & 2.63 & 2.99 \\
\pgmnewton       & 2.21   & 9.49 & 10.21 & 113.75 & 114.02 & 8.67 & 8.84 \\
\pgmprime        & 3.22   & 7.26 & 5.17 & 25.38 & 18.91 & 8.42 & 8.75 \\
\hline
\end{tabular}
\caption{The wallclock runtime (seconds) for each technique per program. 
  }
\label{tbl:evaluationwallclock}
\end{table*}

We evaluate the following mutation testing optimization approaches:
\begin{itemize}
  \item \textbf{Traditional.} The traditional mutation analysis method. Each
    mutants is generated, and executed separately with no sharing.
  \item \textbf{Split-stream.} The split-stream execution is an improvement over
    the traditional mutation execution. Each mutant shares the execution path
    with the original until the mutation is first executed. We use the
    \emph{topsy-turvy}~\cite{gopinath2016topsy} implementation.
  \item \textbf{Modulo-equivalent.} This is an implementation of the \emph{equivalence modulo states}
    approach~\cite{wang2017faster}. This approach is similar to split-stream,
    with the difference that, rather than forking as soon as a mutation is
    executed, the forking is performed when the mutation leads to a unique
    state. This also means that child processes themselves fork again when two
    mutants that initially produced the same state separate at a later point.
  \item \textbf{Execution-taints} The execution taints approach that we
    introduced in this paper. When a control-flow divergence is detected, the
    diverged mutants fork and continue in a child process. Any function calls
    within the mutants are looked up in the memoization cache so that redundancy
    of execution is limited to the function where the mutants diverge.
  \item \textbf{Execution-taints-Nf-Nm.} The execution taints approach that we
    introduced in this paper, but without the memoization, and without forking.
    That is, when a control-flow divergence is detected, the mainline is allowed
    to proceed, and the function is re-executed with the divergent mutant as the
    mainline after the original returns. Further, the function calls are not
    cached by mainline, which means no sharing of child functions by mutants
    after control-flow divergence.
  \item \textbf{Execution-taints-Nf} The execution taints approach, but
    incorporating memoization of function calls.
  \item \textbf{Execution-taints-Nm} The execution taints approach, incorporating
    forking but not memoization of function calls.
\end{itemize}

Given that our implementation is a prototype, a number of implementation
decisions we made are simple costly. For example, using a library based taint
approach imposes a rather large overhead on execution. Similarly, wrapping
expressions in a \<lambda> for evaluation also imposes large overheads. 
These techniques are external to the actual utility of our research. However,
they have an impact on the computational expenditure of the prototype.
That is, it is not informative to compare the actual wall-clock execution time. 
Hence, for evaluation, we instead focus on the actual number of statements
executed by the Python virtual machine. That is, we track the lines executed
by hooking into the python execution using \<pdb.settrace()>. We then tabulate
the lines executed by the mutants and the lines executed by the infrastructure
separately.

The different mutation analysis techniques are evaluated and presented in \Cref{tbl:evaluationcode}.
The infrasttucture support required is given in  \Cref{tbl:evaluationinfra}.

\section{Discussion}
Mutation analysis is the gold standard in evaluating the oracle quality of a
test suite. However, with the advent of automatic test generators such as
fuzzers, the mind share of mutation analysis as the premier tool has steadily
decreased. This is because fuzzers typically generate millions inputs, and
evaluating the fuzzer quality by independently evaluating each input in a mutant
can be impractical. Hence, for making mutation analysis practical for the
emerging world, we need better ways of evaluating mutants.

The interesting fact is that while mutation analysis contains many opportunities
for optimization by eliminating the redundancy in execution, it has remained
rather difficult to achieve it. While recent advances in mutation
analysis~\cite{wang2017faster} manages to eliminate most redundancies in the
pre-divergence phase, it is not yet complete. Indeed, as we show, the mutants can
still share the execution until the control-flow of the mutants diverge.

We note that no previous work has attempted to merge back the mutation execution when
a diverged mutation execution starts following the original program execution
again.
Our technique is the first to achieve this merge-back.

As our results in \Cref{tbl:evaluationcode} show, we can cut down on the redundant
execution by merging back the execution of mutants at the function boundary.
Our technique achieves an average reduction of $16.7\times$ less execution
($T \times 0.06$) when using traditional execution as a baseline (T).
In comparison, the state of the art Modulo-equivalent approach achieves only
$1.47\times$ less execution ($T \times 0.68$) than traditional.

Considering the impact of individual pieces, memoization seems to have an
impact only on \pgmprime. However, we note that this is an artifact of the
small programs we use which do not call other functions from the mutated
function. However, this is not the case in larger programs where most
functions call other functions for various parts of the functionality.
Hence we expect the memoization to make more difference in the real world.

As \Cref{tbl:evaluationinfra} shows, each advancement over mutation analysis
is supported by more complex infrastructure code. For example, for
\pgmcaesarcypher, traditional mutation analysis requires no infrastructure support
during the mutation runtime (the program mutants can simply be executed),
while the execution-taints variants requires the largest support (we note that
in most cases, the infrastructure support required by execution-taints is
comparable to the state-of-the-art modulo-equivalent technique). One thing to
keep in mind here is that the infrastructure code in our implementation is not
optimized, and can likely be optimized heavily in a full-fledged implementation.
For example, both tainting and memoization can have language-runtime support,
which can alleviate their execution cost.

\subsection{Relaxing Forking Constraints}

One of the problems with fork based sharing is that one-shot external resources
cause havoc. To illustrate, consider the following code.
\begin{lstlisting}[style=Python, escapechar=|]
res = create_external_resource(my_resource)
# fork here.
if x > 0: # m1: x > 1
  use_resource(res)
delete_external_resource(res)
\end{lstlisting}
Here, forking happens after the external resource is acquired, and hence both
parent as well as child process uses the resource, and at the end, deletes the
resource. The problem here is that if \<my_resource> can only be used once per
creation, either the parent or the child process will fail for forking but not
for traditional execution.

With our ability to merge the execution back, we can already ensure that the
resource is released only once. Further \<use_resource(res)> is memoized by
the main program, which means that the actual external resource is only used
once by the main.

However, there is a subtlety here. External resources hold their own state,
and can respond differently based on their state. The program has no way to
access the internal state of the external resource. We instead use a different
technique. The idea is to rely on the sequence of calls from the original as
indicating the internal state. We associate a counter to the external resource
such that each operation to the external resource takes the counter as one of
the inputs to the memoization, and the counter is incremented for each call.
Hence, when retrieving the memoized result, we use the corresponding counter
value, along with the rest of the parameters, which gives us back the right result.

If the counter value + parameters is not present in the memoization cache, then
we mark the mutant as wounded. That is, the mutant needs to be run independently
to verify that it has failed.


For dealing with objects, a similar method is useful. The problem with objects
is that the \<self> parameter to any object call contains not just the object id,
but also the entirety of the object state.

\begin{lstlisting}[style=Python, escapechar=|]
obj = MyClass()
# these two are different calls.
obj.add(2)
obj.add(2)
\end{lstlisting}
When an object contains a large number
of members, this can become expensive. Hence, we may associate a counter to each
object that increments when the object is updated, which can be used as a cheap
proxy for the object state.



Further, in this research, we provide several novel ideas, each of which are discussed
below.

\subsection{Repurposing Taints for Mutation Analysis}
Many languages such as Perl~\cite{perlsec} Ruby~\cite{flanagan2008ruby},
JVM and .NET runtime~\cite{livshits2012dynamic,haldar2005dynamic} either provide
the ability to track dynamic data-flow taints, or make it easy and efficient
to implement it. Even if the language in question does not directly support
dynamic taint tracking, one can retrofit it as a
library~\cite{conti2010taint,fuzzingbook2021:InformationFlow}.
Our research is the first to identify that the dynamic taints
can be repurposed for mutation analysis by making mutations induce taints in
the expressions they operate on.
This allows us to avoid redundant computation in mutants that follow the same
control-flow path as the mainline.
Indeed, simply by keeping the mutation results as execution taints can allow us
to go beyond the state of the art (Equivalence Modulo State) by forking only
when the mutants diverge in control-flow rather than when they lead to a new
state.

\subsection{Merge Post-Divergence Execution}
While previous researchers investigated techniques to share pre-divergence
computation, our technique is the first to allow sharing of post-mutation
computation. This shared execution can remove a significant amount of redundant
execution from mutation analysis, making it feasible for advanced test
generators such as fuzzers to use mutation analysis.

\subsection{Memoization}
While memoization to alleviate expensive computations is a well known technique,
our technique shows how to utilize it effectively for mutation analysis. Using
memoization, we are able to effectively limit redundant computation to
intra-function control-flow.

We note that memoization allows us to relax the constraints that were
associated with split-stream and other forking techniques. The main issue can
be demonstrated by considering a test case which involves creating and deleting a file.
Assume that a mutation occurred after creating the file, but before the deletion.
In such case, any forking implementations that fork on the mutation will
fail because a file is created only once, but attempt is made to delete it twice.

With our technique, there are two reasons this can succeed. In the first place,
if the merge back happens before the file deletion, the file will be deleted only
once. Even if the file is deleted before the merge back, the file deletion call
will be memoized, and will be executed only once.


\section{Related Work}
With the advent of automatic test generators, it has become important to ensure
that the test suites generated are actually effective in revealing and
preventing bugs. Mutation analysis is of course the premier tool for that
purpose~\cite{demillo1980mutation}. Mutation analysis involves generating
syntactic variants of the program, which are then evaluated against a test
suite. If the test suite successfully induces a failure in the mutant, but no
failure in the original, we consider the mutant to have been killed. The number
of such mutants killed is called the mutation score of the test suite, and is
considered to be a reliable proxy for the effectiveness of the test suite in
revealing faults in the program. Further, the undetected mutants in a program
for a given test suite is considered a reasonable proxy for the residual defects
in the program.

The error model in mutation analysis is that errors are caused during
transcription of the mental model of the programmer to code~\cite{demillo1978hints}.
This leads to the two axioms of mutation analysis. The \cph{}
and the \ce{}. The \cph{} states that programmers tend to create near-correct
versions of programs, and the \ce{} states that complex faults are coupled to
simple faults such that test cases capable of detecting the simple faults are
also capable of detecting most complex faults.

Combining these pieces, to be effective, mutation analysis needs to exhaustively
generate all possible simple faults, and evaluate them. If the test suite is
able to capture all such simple faults, we can have high confidence that it will
be able to detect all complex faults in the program.

Unfortunately, exhaustively generating and evaluating all simple faults is
costly. Compared to coverage measures, which can be obtained in a single run of
a test suite, mutation analysis (at worst) requires as many runs of the test
suite as there are mutants (because each mutant needs to be independently
evaluated).

Numerous techniques exist to reduce the computational expenditure involved in
traditional mutation analysis~\cite{papadakis2019mutation,pizzoleto2019systematic}.
These can broadly classified into approaches
that try to approximate the traditional mutation score
(weak and firm mutation~\cite{offutt1993experimental,howden1982weak,offutt1991strong,offutt1994empirical,durelli2012toward},
checked coverage~\cite{schuler2013checked}), and those that try to improve the
efficiency or speed of traditional mutation score.


This research focuses on improving the efficiency of evaluation. Hence, we will
take a closer look in this area.
Some of the work in this area involves parallelization of
mutation analysis using MIMD~\cite{offutt1992mutation} and SIMD~\cite{krauser1991high}
machines, in HPC systems~\cite{canizares2016eminent}, and using Hadoop~\cite{saleh2015hadoopmutator}.

Two algorithms try to reduce the number of mutants to be evaluated by
identifying which tests test which mutants. The first called lazy mutation
analysis by Fleyshgakker~\cite{fleyshgakker1994efficient}
which uses weak mutation kills to identify which tests to run on which
mutants. A less stringent approach is to use code coverage for the same purpose~\cite{schuler2009javalanche}.

Mutant schemata~\cite{untch1993mutation} is a technique that encodes all
mutations into a meta mutant. The particular mutant to be executed is passed to
the meta mutant as a command line argument or an
environment variable.
Just et al.~\cite{just2014efficient} show that not all
mutations result in a unique state, and many mutants can be represented by a
single execution.
Split-stream execution~\cite{king1991fortran,gopinath2016topsy,tokumoto2016muvm}
is another technique that tries to reduce the redundancy in mutation executions.
In this technique, the mutants are forked off from the execution only after
encountering the particular mutation in the original. The equivalence modulo
states approach~\cite{wang2017faster} combines partitioning states with
execution forking to further reduce the redundancy in pre-divergence execution.

\section{Threats to Validity}
\subsection{Threats to External Validity}
Threats to external validity are factors that may reduce the generalizability of
our findings. Our approach is implemented only as a prototype. Further, the
subjects were chosen such that they were small, contained operations only on
primitive data types such as \<int> and \<float>. Hence, the results from our
evaluation is not a proof of generality of our technique, rather that the
technique is a promising direction. Indeed, we accept that any number of
uncontrolled parameters such as the subject type, test suites, the mutation
operators used, all can have an impact on the amount of execution saving
accomplished. We propose to study these in detail in future work.

\subsection{Threats to Internal Validity}
Threats to internal validity are factors concerning the judgment of cause and
effect relationships. We have used the number of statements executed by the
Python runtime as an indication of the effectiveness of the technique.
However, this assumes that the supporting infrastructure code can be
optimized such that the actual program execution forms the largest chunk.
While we believe that this is a reasonable assumption, it is possible
that the infrastructure code can't be optimized as heavily as we expect, and
this may lead to a lose in efficiency. This is a threat to validity that we
acknowledge. Further, software bugs are a fact of life, and it is possible
that we may have bugs in our implementation leading to faulty measurements.
This is also another threat that we acknowledge. The mitigation of these
threats will also form our future work.


\section{Limitations and Future Work}
Our implementation is only a prototype. Hence it is subject to a large number of
constraints.
\subsection{Theory}
\begin{itemize}
  \item When control-flow divergence is detected, only mutants in the current
    execution path continue. This means that the computations carried out in
    execution taints of the wounded mutants are discarded.
  \item In a function, after the control-flow divergence, we only merge the
    execution at the function return. However, there could be other parts
    especially after the diverged control-flow structure finishes where
    the execution may be shared. This execution path is not currently shared.
    (This limitation is true only about the execution path within the function.
    As soon as a function call is made from the control-flow with the same
    parameters, the computation is shared across mutants.)
    Even if one merges the execution flow immediately after the control-flow
    structure finishes, there may be parts of control-flow where different
    branches are similar, but do not share the execution.
  \item Shared with split stream and equivalence modulo state: Flow that
    requires single-shot initialization can't be handled.
\end{itemize}

\subsection{Prototype}
\begin{itemize}
  \item Only a limited number of mutation primitives are implemented.
  \item Only integers and floats are implemented.
  \item We do not consider globals.
  \item Multi-threaded programs, programs using fork
  \item Programs using IPC.
\end{itemize}
\todo{Faster Mutation Analysis with Fewer Processes and
Smaller Overheads}

\section{Conclusion}
Mutation analysis is an effective but costly method for evaluating test suite
quality. The main problem is that independent evaluation of individual mutants
involve a large amount of redundancy due to similar pre-divergence and
post-mutation executions.
Previous work explored methods to reduce the pre-divergence execution. However,
post-mutation execution has remained unsolved. We describe a novel method based
on execution taints that can merge post-mutation execution, and hence avoid
post-mutation execution redundancy.



%
\bibliographystyle{IEEEtran}
\bibliography{mutation22-shadow}

\end{document}